\documentclass[aps,twocolumn,nofootinbib,superscriptaddress,pra]{revtex4-2}
\usepackage{graphicx}
\usepackage{amsmath}
\usepackage{amssymb}
\usepackage{amsfonts}
\usepackage{mathtools}
\usepackage{color}
\usepackage{dsfont}
\usepackage{float}
\usepackage{soul}
\usepackage{hyperref}
\usepackage{tikz}
\usetikzlibrary{positioning,shapes}

\definecolor{lime}{HTML}{A6CE39}
\DeclareRobustCommand{\orcidicon}{%
	\begin{tikzpicture}
	\draw[lime, fill=lime] (0,0) 
	circle [radius=0.16] 
	node[white] {{\fontfamily{qag}\selectfont \tiny ID}};
	\draw[white, fill=white] (-0.0625,0.095) 
	circle [radius=0.007];
	\end{tikzpicture}
	\hspace{-2mm}
}
\foreach \x in {A, ..., Z}{%
	\expandafter\xdef\csname orcid\x\endcsname{\noexpand\href{https://orcid.org/\csname orcidauthor\x\endcsname}{\noexpand\orcidicon}}
}

\begin{document}
\title{Quantum simulation in hybrid transmission lines}
\date{\today}

\author{Alessandro Ferreri \orcidA{}}
\affiliation{Institute for Quantum Computing Analytics (PGI-12), Forschungszentrum J\"ulich, 52425 J\"ulich, Germany}
\author{Frank K. Wilhelm \orcidB{}}
\affiliation{Institute for Quantum Computing Analytics (PGI-12), Forschungszentrum J\"ulich, 52425 J\"ulich, Germany}

\begin{abstract}
Platforms based on transmission lines are nowadays employed for the simulation of standard phenomena in quantum electrodynamics and quantum field theory. In this work, we propose a hybrid platform, in which a right-handed transmission line is connected to a left-handed transmission line by means of a superconducting quantum interference device (SQUID). 
We examine the interaction between the two transmission lines, as well as the excitation flow along the composed platform. We show that, by activating specific resonance conditions, this platform can be used as a quantum simulator of different phenomena in quantum optics, multimode quantum systems and quantum thermodynamics.
\end{abstract}

\maketitle

\section{Introduction}
Quantum simulation occupies a preeminent role in the development of state-of-the-art quantum technologies for the study of complex quantum phenomena which could not be investigated via direct observations \cite{RevModPhys.86.153,you_atomic_2011, wilson_photon_2010,wilson_observation_2011,steinhauer2016observation, PhysRevA.69.033602, munoz_de_nova_observation_2019,shi_quantum_2023,e17106893}.
Nowadays, we can account for a vast range of quantum systems that can be employed as quantum simulator. 
Among these systems, particularly successful platforms for implementations in quantum simulation are the superconducting circuits \cite{doi:10.1126/science.1231930,vool_introduction_2017,wendin_quantum_2017}. Depending on the arrangement of the circuit elements, these devices can simulate e.g. two-level systems (qubits), and quantum harmonic oscillators, and for this reason, superconducting circuits find application as quantum simulators of a large number of physical scenarios in quantum electrodynamics (QED) \cite{ blais_circuit_2021, blais_cavity_2004}, atomic physics and quantum optics \cite{you_atomic_2011,weisl_kerr_2015}.

Superconducting electrical circuits are particularly effective for the investigation of multimode quantum systems, such as quantum fields or many-body systems \cite{Zhang2023simulatinggauge}. Typical examples are transmission lines (TLs), which well describe (1+1)-dimensional quantum fields either in free space or confined in a resonator \cite{yurke_quantum_1984}. In the last two decades, the solid analogy between TLs and quantum scalar fields, as well as the possibility to externally drive the magnetic flux by means of superconducting quantum interference devices (SQUID), was of great support for the investigation of quantum relativistic phenomena, with particular focus on those stemming from the stimulation of the quantum vacuum \cite{birrell1984quantum}. For instance, effects of particle creation have been predicted in diverse waveguidelike TLs \cite{nation_colloquium_2012, tian_analog_2017,lang_analog_2019}; well-known examples thereof are the dynamical Casimir effect \cite{lahteenmaki_dynamical_2013,johansson_dynamical_2009, johansson_dynamical_2010} (then observed experimentally \cite{wilson_photon_2010,wilson_observation_2011}), and the Hawking radiation \cite{nation_analogue_2009, tian_analogue_2019, blencowe_analogue_2020}.

Continuous transmission lines are normally right-handed, meaning that the dielectric constant and the magnetic permeability are both always positive. Due to the right-handedness, the dispersion relation of such TLs enables the promotion of the discrete circuit nodes to the continuous limit \cite{blais_circuit_2021}. However, almost seventy years ago Vaselago proposed left-handed media  \cite{veselago1967electrodynamics}, which are characterized by negative dielectric constant and magnetic permeability; moreover, the wave vector points to the opposite direction with respect to the Poynting vector \cite{PhysRevLett.84.4184}. As a consequence of the left-handedness, in a limited range of frequencies such metamaterial also respond to the electromagnetic field with a negative refractive index \cite{PhysRevLett.85.2933, PhysRevLett.90.137401}. 
In circuit QED, circuit platforms showing left-handed features in the dispersion relation are called ``left-handed transmission lines (LHTLs)" \cite{kozyrev_nonlinear_2008,ferreri2024particle}.

In this work, we present a hybrid platform realized by joining a left-handed and a right-handed transmission line (RHTL) \cite{caloz_novel_2004, egger_multimode_2013,messinger_left-handed_2019,wang_mode_2019}. To join the two TLs, we make use of a SQUID. We place this device on the edge between the two TLs, such that it can directly couple the magnetic flux stemming from both TLs. 
We examine the energy conservation at the SQUID, showing that the interaction between the two TLs leads to a energy shift term altering the frequency matching between left-handed and right-handed modes. Importantly, by tuning the Josephson energy of the SQUID we can externally drive the dynamics of the composed transmission line (CTL). We observe that, by properly manipulating the Josephson energy, we can simulate a large spectrum of phenomena described by a linear dynamics (e.g. quantum optics, including nonlinear optics phenomena up to the second order susceptibility). Finally, we propose some applications in quantum thermodynamics.
\begin{figure*}[t]
	\centering
	\includegraphics[width=1\linewidth]{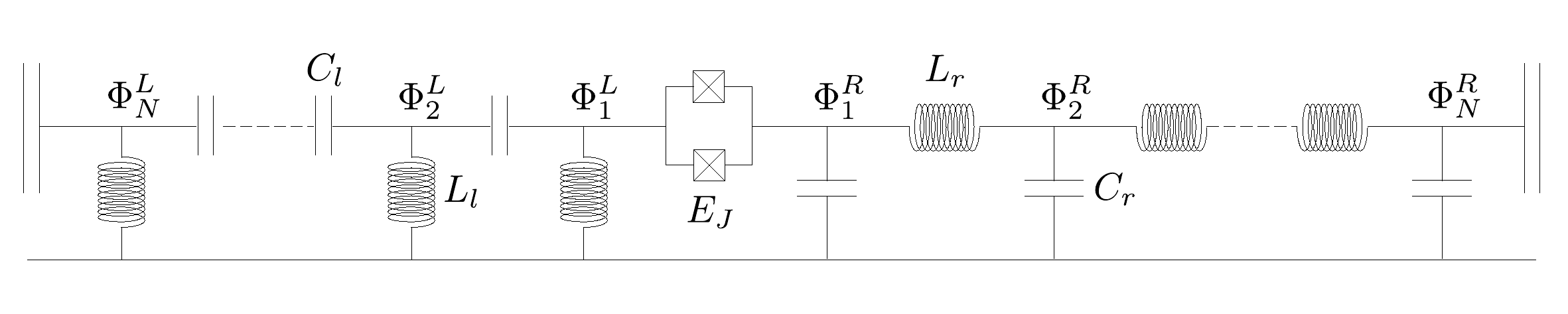}
	\caption{Schematic representation of the hybrid transmission line. It consists of a left-handed and a right-handed transmission line, each of N loops, connected by a SQUID in serie.}
	\label{ibrida}
\end{figure*}

The paper is structured as follows: in Sec.~\ref{model} we present the CTL and discuss the differences between LHTL and RHTL in terms of dispersion relations and eigenenergies of the Hamiltonian. In Sec.~\ref{encon} we address the problem of the energy conservation at the SQUID, and examine the possibility to propagate quanta of magnetic flux fields from one TL to the other one. In Sec.~\ref{intdyn} we show that the interaction Hamiltonian can display different types of two-mode couplings depending on the activated resonance condition between the Josephson energy and the modes of both the LTHL and the RHTL. We also study the correlation between these modes. In Sec.~\ref{sim} we present different scenarios wherein our CTL can find application as quantum simulator and quantum thermal machine. We conclude in Sec.~\ref{conc}. Starting from basic circuit equations, in appendix~\ref{app:hyb} we mathematically achieve the Hamiltonian from the Lagrangian using the magnetic fluxes as scalar fields, and analyze the single terms of the Hamiltonian of interaction.



\section{Theoretical background}\label{model}

The hybrid TL is pictorially represented in Fig.\ref{ibrida}. It is composed of two different platforms, namely a LHTL and a RHTL, connected together by one SQUID at the edge in between. 
We assume that the CTL consists of $2N$ identical unit cells, where $N$ is the number of discrete unit cells equal in both the LHTL and RHTL. Each unit cell is characterized by the length unit $\Delta x$, determining the minimum wave length for the transmission lines, $\lambda=2\Delta x$, and confining the wave vector within the first Brillouin zone \cite{egger_multimode_2013}.
The presence of the SQUID with the time-dependent Josephson energy ensures the interaction between the two TLs; furthermore, we will see that it allows to impose the energy conservation for selected modes of the CTL.

The Lagrangian of the hybrid transmission line consists of three parts, two thereof describe the left-handed and the right-handed component individually, whereas the last term is responsible for their interaction:
\begin{align}
\mathcal{L}=\mathcal{L}_\textrm{L}+\mathcal{L}_\textrm{R}+\mathcal{L}_\textrm{I}
\label{Lagrangian}
\end{align}
where
\begin{subequations}\label{Lterms}
\begin{align}
\mathcal{L}_\textrm{L}=&\frac{1}{2}\sum_{n_\textrm{l}=1}^N\left[C_\textrm{l}(\dot\Phi_{n_\textrm{l}+1}^\textrm{L}-\dot\Phi_{n_\textrm{l}}^\textrm{L})^2-\frac{(\Phi_{n_\textrm{l}}^\textrm{L})^2}{L_\textrm{l}}\right],\label{LL}\\
\mathcal{L}_\textrm{R}=&\frac{1}{2}\sum_{n_\textrm{r}=1}^N\left[C_\textrm{r}(\dot\Phi_{n_\textrm{r}}^\textrm{R})^2-\frac{(\Phi_{n_\textrm{r}+1}^\textrm{R}-\Phi_{n_\textrm{r}}^\textrm{R})^2}{L_\textrm{r}}\right],\label{LR}\\
\mathcal{L}_\textrm{I}=&E(t)\cos\left[\frac{2\pi}{\phi_0}\left(\Phi_1^\textrm{L}-\Phi_1^\textrm{R}\right)\right].
\end{align}
\end{subequations}
In the expressions above, $\Phi_{n_\textrm{l}}^\textrm{L}$ ($\Phi_{n_\textrm{r}}^\textrm{R}$) represents the value of the magnetic flux field of the left-handed (right-handed) transmission line at the position $n_\textrm{l}$ ($n_\textrm{r}$), $C_\textrm{l}$ and $L_\textrm{l}$ are the capacitance and the inductance of the LHTL, whereas $C_\textrm{r}$ and $L_\textrm{r}$ are the capacitance and inductance of the RHTL, respectively. Note that the higher $n_\textrm{l}$ and $n_\textrm{r}$, the farther the labelled position with respect to the SQUID. Finally, $E(t)$ is the time-dependent Josephson energy, and $\phi_0=\pi\hbar/e$, is the magnetic flux quantum.

For the sake of simplicity, in our description we assume that the capacitance of the Josephson junctions in the SQUID is small with respect to the capacitance at both the LHTL and the RHTL, $C_\textrm{J}\ll C_\textrm{l},C_\textrm{r}$, implying that we can neglect it in our description. 
Importantly, we will work in a regime where $\Phi_0/\phi_0\ll 1$ \cite{johansson_dynamical_2009,johansson_dynamical_2010, weisl_kerr_2015, tian_analog_2017}, thereby writing the linearized interaction Lagrangian as
\begin{align}
    \mathcal{L}_\textrm{I}\simeq&-\frac{E(t)}{2}\left(\frac{2\pi}{\phi_0}\right)^2\left(\Phi_1^\textrm{L}-\Phi_1^\textrm{R}\right)^2.
\label{Lint}
\end{align}
In Eq.\eqref{Lint} we expanded the cosine to the lowest order in $2\pi\left(\Phi_1^\textrm{L}-\Phi_1^\textrm{R}\right)/\phi_0$.

Solving the Euler-Lagrange equations far from the SQUID, namely at $1<n_\textrm{l},n_\textrm{r}<N$, we get the discrete mode expansions of the quantum flux field in both the LHTL and the RHTL in terms of plane wave:
{\small
\begin{align}
\hat\Phi^\textrm{L}(n_\textrm{l},t)=&\sum_{\lvert j \rvert=1}^{N/2}\sqrt{\frac{\hbar}{2C_\textrm{l} N\omega_{j}}}\left(e^{i(k_j n_\textrm{l}\Delta x-\omega_{j} t)}a_j+\textrm{h.c.}\right),
\label{leftmode}\\
\hat\Phi^\textrm{R}(n_\textrm{r},t)=&\sum_{\lvert j \rvert=1}^{N/2}\sqrt{\frac{\hbar}{2C_\textrm{r} N \upsilon_{j}}}\left(e^{i(p_j n_\textrm{r}\Delta x- \upsilon_{j} t)} b_j+\textrm{h.c.}\right),
\label{rightmode}
\end{align}
}
where $\omega_j$ and $k_j$ are frequency and wave vector of the left-handed TL, respectively; whereas $\upsilon_j$ and $p_j$ are frequency and wave vector of right-handed TL, respectively. The sum over all modes accounts for plane wave propagating both leftwards (negative sign of $j$) and rightwards (positive sign of $j$) along the CTL. Here, $a_j$ and $b_j$ are the classical amplitudes of the LHTL and RHTL magnetic flux fields respectively.
By means of the Euler-Lagrange equations, we also find that the two TLs are characterized by the following dispersion relations
\begin{align}
\omega_j=&\frac{1}{2\sqrt{C_\textrm{l}L_\textrm{l}}\left\lvert\sin\left(\frac{k_j\Delta x}{2}\right)\right\rvert}, \label{displeft}\\
\upsilon_j=&\frac{2\left\lvert\sin\left(\frac{p_j\Delta x}{2}\right)\right\rvert}{\sqrt{C_\textrm{r}L_\textrm{r}}}, \label{dispright}
\end{align}
for the LHTL and RHTL, respectively.
Note that the form of the wave vectors is identical for both TLs, $w_j=\frac{2\pi j}{N\Delta x}$, with $w=k,p$. Referring to Fig.\ref{ibrida}, in our notation the positive sign of the wave vectors always indicates the rightwards propagation of the signal.

The quantization of the two classical magnetic flux fields occurs in accord to the standard procedure \cite{louisell1973quantum, ferreri2024particle}: we first promote the classical fields in Eq.\eqref{leftmode} and Eq.\eqref{rightmode}, as well as the canonical momenta $P^\textrm{L}(n_\textrm{l},t)$ and $P^\textrm{R}(n_\textrm{r},t)$ defined respectively in Eq.\eqref{momL} and Eq.\eqref{momR}, to quantum operators: these fulfill the bosonic commutation relation 
\begin{align}
\left[\hat\Phi^\textrm{L}(n,t),\hat P^\textrm{L}(m,t)\right]=\left[\hat\Phi^\textrm{R}(n,t),\hat P^\textrm{R}(m,t)\right]=i\hbar\delta_{nm},
\label{commPP}
\end{align}
whereas all other commutators are equal to zero. The quantization of the classical amplitudes is derived by discrete Fourier transforming Eqs.\eqref{leftmode}, \eqref{rightmode}, \eqref{momL} and \eqref{momR} and exploiting the commutation rule in Eq.\eqref{commPP} \cite{ferreri2024particle}. We get
\begin{align}
\left[\hat a_i,\hat a_j^\dag\right]=\frac{\delta_{ij}}{4\sin^2\left(\frac{k_j\Delta x}{2}\right)},
\label{commaad}
\end{align}
for the magnetic flux field in the LHTL and
\begin{align}
\left[\hat b_i,\hat b_j^\dag\right]=\delta_{ij},
\label{commbbd}
\end{align}
for the magnetic flux field in the RHTL.
With little abuse of notation, sometimes we will refer to the quantum excitations of the magnetic flux fields as ``photons".

Starting from the Lagrangian in Eq.\eqref{Lagrangian} and the explicit expressions for the two magnetic flux fields written above, we can derive the Hamiltonian $\mathcal{\hat H}=\mathcal{\hat H}_\textrm{L}+\mathcal{\hat H}_\textrm{R}+\mathcal{\hat H}_\textrm{I}$, where
\begin{align}
\mathcal{\hat H}_\textrm{L}=&4\hbar\sum_{\lvert j \rvert=1}^{N/2}\omega_j\sin^2\left(\frac{k_j\Delta x}{2}\right)\hat a_j^\dag\hat a_j \label{HL}\\
\mathcal{\hat H}_\textrm{R}=&\hbar\sum_{\lvert j \rvert=1}^{N/2} \upsilon_j\hat b_j^\dag\hat b_j\label{HR}
\end{align}
constitute the two parts of the noninteracting Hamiltonian for the left-handed and right-handed transmission lines, $\mathcal{\hat H}_0=\mathcal{\hat H}_\textrm{L}+\mathcal{\hat H}_\textrm{R}$, and 
\begin{align}
\mathcal{\hat H}_\textrm{I}=\chi(t)&\left\{\sum_{\lvert j \rvert=1}^{N/2}\left[\frac{1}{\sqrt{C_\textrm{l}\omega_j}}\left(e^{-ik_j \Delta x}\hat a_j^\dag+e^{ik_j \Delta x}\hat a_j\right)\right.\right.\nonumber\\
&\left.\left.-\frac{1}{\sqrt{C_\textrm{r}\upsilon_j}}\left(e^{-ip_j\Delta x}\hat b_j^\dag+e^{ip_j\Delta x}\hat b_j\right)\right]\right\}^2
\label{intHamold}
\end{align}
with $\chi(t)=\frac{\hbar E(t)}{2N }\left(\frac{2\pi}{\phi_0}\right)^2$,
is the interaction Hamiltonian. The mathematical derivation of the Hamiltonian is reported in Appendix \ref{app:hyb}. We notice that, as long as we do not account for the interaction at the SQUID, the eigenenergies in Eq.\eqref{HL} are structurally identical to the eigenvalues (frequencies) in Eq.\eqref{HR}, 
\begin{align}
 \epsilon_j\equiv 4\hbar\omega_j\sin^2\left(\frac{k_j\Delta x}{2}\right)=\frac{2\hbar\left\lvert\sin\left(\frac{k_j\Delta x}{2}\right)\right\rvert}{\sqrt{C_\textrm{l}L_\textrm{l}}}.   
\end{align}
Therefore, the bare energy grows with the wave vector up to the border of the first Brillouin zone in both LHTL and RHTL, despite the different structure of the dispersion relations \cite{ferreri2024particle}.

\section{Phase matching at the SQUID}\label{encon}
In this section we examine the signal propagation along the CTL. In particular, we want to figure out whether the interaction between two TLs characterized by different dispersion relations enables energy and momentum propagation at their intersection.
To address this issue properly, we need to analyze the interaction Hamiltonian in detail. 

Once we expand the square in Eq.\eqref{intHamold}, we observe that the interaction Hamiltonian can be split conveniently into smaller terms, each representing a specific type of two-mode coupling:
\begin{align}
\mathcal{\hat H}_\textrm{I}=\mathcal{\hat H}_\textrm{ES}+\mathcal{\hat H}_\textrm{HP}+\mathcal{\hat H}_\textrm{RM}+\mathcal{\hat H}_\textrm{1S}+\mathcal{\hat H}_\textrm{2S}+\mathcal{\hat H}_\textrm{IS}
\label{intHam}
\end{align}
More details about the single contributions of the Hamiltonian in Eq.\eqref{intHam} and their explicit form are discussed in Appendix \ref{app:hyb}. Note that the interaction can occur both within the same TL, and between modes stemming from the left-handed and right-handed components.

We now need to figure out whether all terms of the interaction Hamiltonian contribute to the dynamics with equal weigh, or whether there are resonant terms whose contribution play a major role. To verify this, we momentarily move our description to the interaction picture, and observe if there are static terms in the Hamiltonian. 
To induce an external time-dependence in the Hamiltonian at $t>0$ we modulate the Josephson energy of the SQUID around a fixed value according to $E(t)=E_0[1+\epsilon\cos(\Omega t)]$, where $E_0=I_\textrm{c} \phi_0$ is the average value of the Josephson energy with critical current $I_\textrm{c}$. Here, $\Omega$ and $\epsilon$ are the oscillation frequency and the dimensionless oscillation amplitude, respectively.

Once we move to the interaction picture, the only constant term in the interaction Hamiltonian is
\begin{align}
\mathcal{\hat H}_\textrm{ES}^\textrm{I}=\frac{\hbar E_0}{N}\left(\frac{2\pi}{\phi_0}\right)^2\sum_{\lvert j \rvert=1}^{N/2}\left(\frac{1}{C_\textrm{l}\omega_j}a_j^\dag\hat a_j+\frac{1}{C_\textrm{r}\upsilon_j}\hat b_j^\dag\hat b_j\right),
\end{align}
which describes the shift of the bare energies and frequencies in both transmission lines. We can therefore redefine the noninteracting Hamiltonian by including this term as $\mathcal{\hat H}_0=\mathcal{\hat H}_\textrm{L}+\mathcal{\hat H}_\textrm{R}+\mathcal{\hat H}_\textrm{ES}$. The energy shift of the bare eigenenergies in the LHTL is simply given by:
\begin{align}\label{epsilon}
\tilde\epsilon_j=\epsilon_j+\frac{\hbar E_0}{N C_\textrm{l}\omega_j}\left(\frac{2\pi}{\phi_0}\right)^2.
\end{align}
To see the frequency shift clearly, we now introduce the corrected dispersion relations of the LHTL and the RHTL as follows:
\begin{align}
\tilde\omega_j=&\omega_j+\frac{ E_0}{4N C_\textrm{l}\omega_j\sin^2\left(\frac{k_j\Delta x}{2}\right)}\left(\frac{2\pi}{\phi_0}\right)^2,\label{lhtlw}\\
\tilde\upsilon_j=&\upsilon_j+\frac{E_0}{N C_\textrm{r}\upsilon_j}\left(\frac{2\pi}{\phi_0}\right)^2.\label{rhtlw}
\end{align}
These are achieved by letting the ladder operators evolve in the interaction picture, and exploiting the commutation rules in Eqs.\eqref{commaad} and \eqref{commbbd}:
\begin{align}
\hat a_j^\textrm{I}(t)=\hat U_0^\dag(t)\hat a_j
\hat U_0(t)=\hat a_j e^{-i\tilde\omega_j t},\\
\hat b_j^\textrm{I}(t)=\hat U_0^\dag(t)\hat b_j
\hat U_0(t)=\hat b_j e^{-i\tilde\upsilon_j t},
\end{align}
where $\hat U_0=e^{-i\mathcal{H}_0t/\hbar}$.
Each corrected frequency consists of the sum of the bare frequency and a correcting term. Interestingly, in both LHTL and RHTL such correcting terms decrease monotonically with the wave vector.

Now we want to focus on the possibility to transmit excitations through the SQUID. We notice that the only element in the interaction Hamiltonian allowing the exchange of excitations between the two parts of the CTL is
\begin{align}
\mathcal{\hat H}_\textrm{HP}^\textrm{I}=\sum_{i,j}^{N/2}g_{ij}&\left(\hat a_i\hat b_j^\dag e^{-i(\tilde\omega_i-\tilde\upsilon_j)t+i(k_i-p_j)\Delta x}+\textrm{h.c.}\right).
\label{HP}
\end{align}
where we defined the coupling constants $g_{ij}=-\frac{\hbar E_0}{N\sqrt{C_\textrm{l}C_\textrm{r}\omega_i\upsilon_j}}\left(\frac{2\pi}{\phi_0}\right)^2\nonumber$.
This term describes the hopping effect between the two TLs, namely the annihilation of one excitation on one side of the CTL and the creation of another excitation on the other side. 
The hopping effect between two modes of the two TLs can be activated by imposing the energy and momentum conservation of the two magnetic flux fields at the SQUID. However, as a consequence of the discreteness of the mode structure, the propagation of quantum excitations through the SQUID does not result possible for all mode pairs, but only for those modes of the magnetic flux fields fulfilling the phase matching conditions given by $k_i=p_j$ (momentum conservation) and the transcendental equation $\tilde\omega_i=\tilde\upsilon_j$ (frequency conservation). Since the momentum conservation of process is fulfilled by modes with the same mode number, the transcendental equation becomes $\tilde\omega_{j}=\tilde\upsilon_{j}$, and it is solved by tuning the set of parameters $\{C_\textrm{l},C_\textrm{r},L_\textrm{l},L_\textrm{r}\}$ opportunely. The discrete mode structure of both TLs therefore allows to restrict the propagation of excitations between the two TLs to only two selected modes, thereby limiting undesired hopping effects between other modes.

\begin{figure}[ht!]
	\centering	\includegraphics[width=1\linewidth]{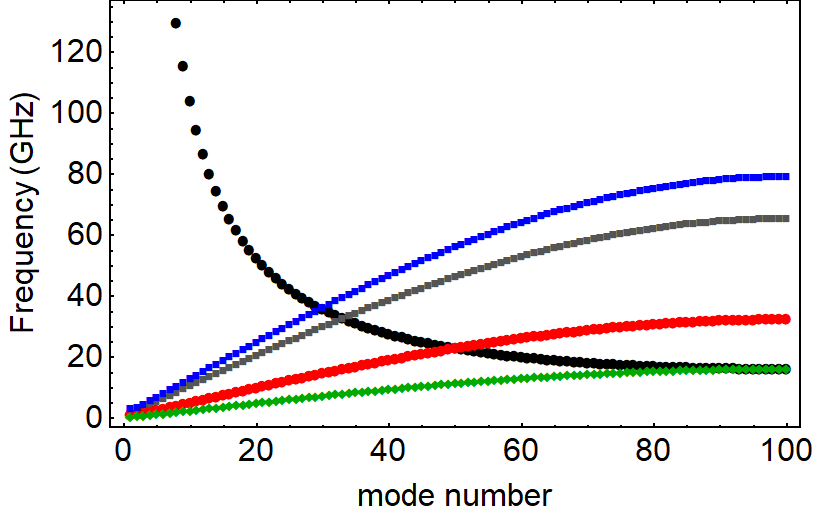}
\caption{Corrected dispersion relations $\tilde\omega_j$ (black) and bare eigenenergies $\tilde\epsilon_j/\hbar$ (gray) of the LHTL. Other colors indicate the corrected dispersion relations $\tilde\upsilon_j$ of the RHTL, achieved by solving the transcendental equation $\tilde\omega_j=\tilde\upsilon_j$, with $j$ labeling the $j$-th mode of both the LHTL and the RHTL, and using $C_\textrm{r}$ as the variable. Frequencies of the RHTL are plotted ensuring the frequency matching for the mode pair $\tilde\omega_{30}=\tilde\upsilon_{30}$ (blue), $\tilde\omega_{50}=\tilde\upsilon_{50}$ (red), and $\tilde\omega_{100}=\tilde\upsilon_{100}$ (green), thereby obtaining $C_\textrm{r}=0.27$ pF, $C_\textrm{r}=1.60$ pF, and $C_\textrm{r}=6.39$ pF, respectively.
Other parameters are: $C_\textrm{l}=0.4$ pF, $L_\textrm{r}=L_\textrm{l}=60$ pH, $I_\textrm{c}=1.25$ $\mu$A, and $N=200$.}
	\label{disp}
\end{figure}

To make a concrete example, in Fig.\ref{disp} we plot the LHTL eigenenergies in Fig.\eqref{epsilon}, as well as the dispersion relations in Eq.\eqref{lhtlw} and Eq.\eqref{rhtlw}. The two dispersion relations are obtained by selecting a set of parameters $\{C_\textrm{l},L_\textrm{l},L_\textrm{r}\}$, and fixing a determined resonant mode pair by solving the trascendental equation $\tilde\omega_{j}=\tilde\upsilon_{j}$ with respect to the capacitance of the RHTL, $C_\textrm{r}$. We solved this equation for three set of modes, namely $\tilde\omega_{100}=\tilde\upsilon_{100}$, $\tilde\omega_{50}=\tilde\upsilon_{50}$ and $\tilde\omega_{30}=\tilde\upsilon_{30}$, obtaining three different values of $C_\textrm{r}$, then we plotted the dispersion relations.


\section{Interaction and dynamics}\label{intdyn}
In the last section we observed that we can control the set of parameters $\{C_\textrm{l},C_\textrm{r},L_\textrm{l},L_\textrm{r}\}$ to enable the propagation of quanta of the magnetic flux fields between two modes of the LHTL and the RHTL. All other terms in Eq.\eqref{HP} describing the interaction between two off-resonant modes do not fulfill the phase matching conditions, and their contribution is in fact negligible. This allows us to drastically reduce the Hilbert space of the system, thereby focusing on few modes of the CTL. 

The possibility to isolate the contribution of specific mode pairs and ignore all secularities (i.e. off-resonant terms) can be applied to other interaction terms accounted in the Hamiltonian in Eq.\eqref{intHam}.
In particular, since we are interested in the interaction between the LHTL and the RHTL, we can tailor the set $\{\Omega, C_\textrm{l}, C_\textrm{r},L_\textrm{l},L_\textrm{r}\}$, thereby ensuring the phase matching for only one coupling term in Eq.\eqref{intHam} containing products of ladder operators stemming from different TLs. Note that the presence of the modulation frequency $\Omega$ is fundamental for the activation of Raman coupling or squeezing terms, which otherwise would be counterrotating and therefore negligible after the rotating wave approximation (RWA).

The possibility to control the Hamiltonian in Eq.\eqref{intHam} and activate one of the possible interaction terms according to our needs makes this device an interesting platform for the simulation of typical scenarios in quantum optics and quantum information.
In what follows, we present some possible application of our findings. We stress that, although we might consider also the coupling between modes within the same TL, henceforth we will focus on the interaction of modes stemming from the LHTL and the RHTL. We will always assume that the two resonant modes are characterized by the same mode number $j$, thereby omitting the label $j$ henceforth.
\subsection{Photon hopping}\label{sec:hp}
As a first scenario, we again consider the interaction
\begin{align}
\mathcal{\hat H}_\textrm{I}=\hbar \xi_\textrm{hp} \left(\hat a^\dag\hat b+\hat a\hat b^\dag\right),
\label{hhop}
\end{align}
with coupling constant
\begin{align}
\xi_\textrm{hp}=-\frac{2E_0}{N\omega\sqrt{C_\textrm{r}C_\textrm{l}}}\left(\frac{2\pi}{\phi_0}\right)^2.
\end{align}
This Hamiltonian can be achieved from Eq.\eqref{HP} by tailoring the parameters such that $\tilde\omega=\tilde\upsilon=\omega$ and performing the RWA. Note that to achieve this result we did not require the modulation of the Josephson energy at the SQUID.
In the Heisenberg picture, the annihilation operators for the two TLs evolve as follows:
\begin{subequations}\label{outputhp}
\begin{align}
\hat a(t)=&e^{-i\omega t}\left(\hat a\cos(\xi_\textrm{hp} t)+i\hat b\sin(\xi_\textrm{hp} t)\right),\\
\hat b(t)=&e^{-i\omega t}\left(\hat b\cos(\xi_\textrm{hp} t)+i\hat a\sin(\xi_\textrm{hp} t)\right).
\end{align}
\end{subequations}
This Hamiltonian describes a collinear beam splitting, in which the transmission and reflection parameters are modulated in time by the coupling constant $\xi_\textrm{hp}$. This type of interaction therefore expresses the propagation of excitations along the CTL due to the annihilation of one photon on one of the two TLs and the creation of one photon with same frequency and momentum on the other one.
\subsection{Raman scattering}\label{sec:rm}
A second interesting scenario is delineated by the Hamiltonian
\begin{align}
\mathcal{\hat H}_\textrm{I}=\hbar \xi_\textrm{rm} \left(\hat a^\dag\hat b+\hat a\hat b^\dag\right),
\end{align}
which is structurally identical to Eq.\eqref{hhop} but with coupling constant
\begin{align}
\xi_\textrm{rm}=-\frac{\epsilon E_0}{N\sqrt{C_\textrm{r}C_\textrm{l}\omega\upsilon}}\left(\frac{2\pi}{\phi_0}\right)^2.
\label{rmxi}
\end{align}
To obtain this Hamiltonian, we tune the SQUID modulation frequency such that $\tilde\omega=\tilde\upsilon+\Omega$, assuming that $\tilde\omega>\tilde\upsilon$.
In contrast to the previous case, the frequencies of the two interacting modes differ, and the frequency conservation is only guaranteed by the modulation of the Josephson energy at the SQUID. This interaction gives rise to two possible phenomena: the Stokes Raman scattering, when the SQUID induces the decay of a photon from the LHTL into a lower-energetic photon of the RHTL; the anti-Stokes Raman scattering, when the SQUID pumps the mode $\tilde\upsilon$ of the RHTL, which decays into a photon $\tilde\omega$ with higher energy. Clearly, we can manipulate our parameters thereby activating a resonance of the form $\tilde\upsilon=\tilde\omega+\Omega$, with $\tilde\upsilon>\tilde\omega$. 

One of the remarkable facts stemming from combining two TLs with different dispersion relations is that the Raman effect can also occur between modes of the two TLs having same bare energy $\tilde\epsilon=\hbar\tilde\upsilon$. After the Raman scattering in the SQUID, a signal generated in the RHTL would therefore propagate to the LHTL with amplified frequency but same wave vector and bare energy \cite{1016358}.

As before, we report the annihilation operators of the two modes $\tilde\omega$ and $\tilde\upsilon$ evolved in the Heisenberg picture:
\begin{subequations}\label{outputrm}
\begin{align}
\hat a(t)=&e^{-i\tilde\omega t}\left(\hat a\cos(\xi_\textrm{rm} t)+i\hat b\sin(\xi_\textrm{rm} t)\right),\\
\hat b(t)=&e^{-i\tilde\upsilon t}\left(\hat b\cos(\xi_\textrm{rm} t)+i\hat a\sin(\xi_\textrm{rm} t)\right).
\end{align}
\end{subequations}
Despite the evident similarities between the annihilation operators written above and those in Eq.\eqref{outputhp}, we stress that this interaction mixes two input modes with different frequencies.
\subsection{Two-mode squeezing}\label{sec:sq}
The modulation of the SQUID can give rise to another interesting scenario. Indeed, by activating the resonance condition $\Omega=\tilde\omega+\tilde\upsilon$ we can squeeze the state of two modes stemming from two different TLs.
Once we fix the resonance condition written above, ensure the momentum conservation $p=-k$, and apply the RWA, the Hamiltonian describing the two-mode squeezing coupling reads
\begin{align}
\mathcal{\hat H}_\textrm{I}=\hbar \xi_\textrm{sq}\left(\hat a^\dag\hat b^\dag+\hat a\hat b\right),
\label{hsq}
\end{align}
and the annihilation operators in the Heisenberg picture evolve as
\begin{subequations}\label{outputsq}
\begin{align}
\hat a(t)=&e^{-i\tilde\omega t}\left(\hat a\cosh(\xi_\textrm{sq} t)-i\hat b^\dag\sinh(\xi_\textrm{sq} t)\right),\\
\hat b(t)=&e^{-i\tilde\upsilon t}\left(\hat b\cosh(\xi_\textrm{sq} t)-i\hat a^\dag\sinh(\xi_\textrm{sq} t)\right),
\end{align}
\end{subequations}
with squeezing parameter $\xi_\textrm{sq}$ equal to $\xi_\textrm{rm}$ in Eq.\eqref{rmxi}. 
These operators suggest us that we can squeeze the quantum vacuum of the two transmission lines at the same time, thereby generating one photon in the LHTL and one photon in the RHTL.
Phenomena of this kind play a crucial role in nonlinear optics. An example is the parametric down-conversion, in which a pump laser interacts with a nonlinear material generating two entangled and spatially distinguishable photons, named signal and idler photons \cite{wu_generation_1986,rarity_observation_1987,PhysRevA.50.5122,Christ_2013}. 

\section{Correlations between the TLs}
In this section we study the correlations between two modes of the LHTL and the RHTL interacting resonantly.  
Our approach is based on the second order correlation function, which, for our purpose, assumes the following expression \cite{scully1999quantum}:
\begin{widetext}
\begin{align}
G^{(2)}(n_\textrm{l},t_1;n_\textrm{r},t_2)=\langle\Phi^{\textrm{L}(-)}(n_\textrm{l},t_1)\Phi^{\textrm{R}(-)}(n_\textrm{r},t_2)\Phi^{\textrm{L}(+)}(n_\textrm{l},t_1)\Phi^{\textrm{R}(+)}(n_\textrm{r},t_2)\rangle.
\label{socf}
\end{align}
\end{widetext}
Sometimes we will make use of the normalized second order correlation function,
\begin{align}
g^{(2)}(n_\textrm{l},t_1;n_\textrm{r},t_2)=\frac{G^{(2)}(n_\textrm{l},t_1;n_\textrm{r},t_2)}{G^{(1)}(n_\textrm{l},t_1)G^{(1)}(n_\textrm{r},t_2)},
\end{align}
where $G^{(1)}(n,t)=\langle\Phi^{(-)}(n,t)\Phi^{(+)}(n,t)\rangle$ is the first order correlation function.

We initialize the system in the two-mode Fock state $\lvert\psi(0)\rangle=\lvert s_\textrm{L}, s_\textrm{R}\rangle$, where $s_\textrm{L}$ and $s_\textrm{R}$ represent the number of excitations of the modes with frequencies $\tilde\omega$ and $\tilde\upsilon$, respectively. This reduces the correlation function to
\begin{align}\label{Gstart}
G_{s_\textrm{L}, s_\textrm{R}}^{(2)}(t_1,t_2)
=\frac{\hbar^2}{N^2C_\textrm{r} C_\textrm{l}\omega \upsilon}\langle\hat a^\dag(t_1)\hat b^\dag(t_2)\hat a(t_1)\hat b(t_2)\rangle.
\end{align}
The explicit form of the correlation function depends on the specific resonance we are interested in. In what follows, we will analyse the three cases studied before, namely the photon hopping, the Raman scattering and the two-mode squeezing. Note that, despite the different physical interpretation of the two scenarios in sections \ref{sec:hp} and \ref{sec:rm}, the annihilation (and therefore also the creation) operators in Eq.\eqref{outputhp} and Eq.\eqref{outputrm} lead to the same form of the correlation function, therefore these two cases will be presented together.
\subsection{Photon hopping and Raman scattering}
We now fix the parameters of the CTL such that the energy conservation is fulfilled either in terms of $\tilde\omega=\tilde\upsilon$, or
$\tilde\omega=\tilde\upsilon+\Omega$. As mentioned above, the second order correlation function is structurally identical for the two scenarios, and reads
\begin{align}
G_{s_h,s_d}^{(2)}(t_1,t_2)=&\frac{\hbar^2 s_\textrm{L} s_\textrm{R}C_1^2C_2^2}{N^2C_\textrm{r} C_\textrm{l}\omega \upsilon}\nonumber\\
&\left[(1-T_1T_2)^2+\frac{s_\textrm{L}-1}{s_\textrm{R}}T_2^2+\frac{s_\textrm{R}-1}{s_\textrm{L}}T_1^2\right],
\end{align}    
where we defined $C_j=\cos(\xi t_j)$ and $T_j=\tan(\xi t_j)$, with $\lvert j \rvert=1,2$.
Here, $\xi$ corresponds to either $\xi_\textrm{hp}$ or $\xi_\textrm{rm}$. Clearly, when the CTL is prepared in the vacuum state, or when only one photon populates the CTL, the correlation function is identically zero. However, it is particularly interesting to consider the case when the CTL is initially populated by two photons, each stemming from a different TL. In this case, the normalized second order correlation function becomes 
\begin{align}
g_{1_\textrm{L}, 1_\textrm{R}}^{(2)}(t_1,t_2)=&\cos^2[\xi(t_1+t_2)].
\label{ncf}
\end{align}
This correlation function tells us that the coincidence probability to detect two photons at different TLs is modulated, and that periodically this probability reaches zero.
To see what happens to the two photons when $g_{1_\textrm{L}, 1_\textrm{R}}^{(2)}(t_1,t_2)=0$, we let the system evolve in the Schrödinger picture. Since the initial state consists of two photons in different channels, this is equivalent to applying the creation operators $\hat a^\dag(t)$ and $\hat b^\dag(t)$ in Eq.\eqref{outputhp} or Eq.\eqref{outputrm} to the vacuum state, $\lvert\psi(t)\rangle= \hat a^\dag(t)\hat b^\dag(t)\lvert 0_\textrm{L}, 0_\textrm{R}\rangle$, hence obtaining 
\begin{align}
\lvert\psi(t)\rangle=& e^{i(\tilde\omega+\tilde\upsilon) t}\cos\left(2\xi\,t\right)\lvert 1_\textrm{L}, 1_\textrm{R}\rangle\nonumber\\
&+i \,e^{i(\tilde\omega+\tilde\upsilon) t}\sin\left(2\xi\,t\right)\left(\lvert 0_\textrm{L}, 2_\textrm{R}\rangle+\lvert 2_\textrm{L}, 0_\textrm{R}\rangle\right).
\label{state}
\end{align}
As we can see, the normalized correlation function in Eq.\eqref{ncf} with $t_1=t_2=t$ corresponds to the square of the first term of the output state, which describes the probability to detect two photons in two different modes. When this probability vanishes at time $t_\textrm{dip}=\pi/(4\xi)$, the probability to find two photons at the same TL is maximized. 
The state $\lvert\psi(t_\textrm{dip})\rangle$ therefore corresponds to the output state of a quantum interference scenario, in which two photons interacting in a beam splitter are only found both in one of the two output channels. 
This phenomenon, in which the quantum interference of two photons annuls the coincidence probability and generates the two-photon NOON state, is referred to as Hong-Ou-Mandel effect, and it finds applications in many quantum optical frameworks \cite{PhysRevLett.59.2044,mattle_dense_1996,bouchard_two-photon_2021,PhysRevA.104.043707}.
\subsection{Two-mode squeezing}
To conclude this section, we analyze the second order correlation function when the resonance $\Omega=\tilde\omega+\tilde\upsilon$ is activated. Substituting Eq.\eqref{outputsq} into Eq.\eqref{Gstart} we obtain
\begin{widetext}
\begin{align}
G_{s_h,s_d}^{(2)}(t_1,t_2)=&\frac{\hbar^2 s_\textrm{L}s_\textrm{R}\bar C_1^2\bar C_2^2}{N^2C_\textrm{r} C_\textrm{l}\omega \upsilon}\left[1+\bar T_1\bar T_2+\frac{(s_\textrm{L}+1)(s_\textrm{R}+1)}{s_\textrm{L}s_\textrm{R}}(\bar T_1^2\bar T_2^2+\bar T_1\bar T_2)+\frac{s_\textrm{L}+1}{s_\textrm{R}}\bar T_2^2+\frac{s_\textrm{R}+1}{s_\textrm{L}}\bar T_1^2\right]
\end{align}
\end{widetext}
where we defined $\bar C_j=\cosh(\xi_\textrm{sq} t_j)$ and $\bar T_j=\tanh(\xi_\textrm{sq} t_j)$, with $\lvert j \rvert=1,2$. This function describes the correlations between the LHTL and the RHTL when the two modes of interest are prepared in a Fock state with $s_\textrm{L}$ and $s_\textrm{R}$ excitations, respectively. As expected for squeezed states, the second order correlation function does not vanish when the system is prepared in the vacuum state. Indeed in this case the normalized second order correlation function becomes
\begin{align}
g_{0_\textrm{L},0_\textrm{R}}^{(2)}(t_1,t_2)=&1+\coth(\xi_\textrm{sq} t_1)\coth(\xi_\textrm{sq} t_2),
\end{align}
and reduces to
\begin{align}
g_{0_\textrm{L},0_\textrm{R}}^{(2)}(t)=&1+\coth^2(\xi_\textrm{sq} t),
\end{align}
when $t_1=t_2$, demonstrating the presence of correlations even when the system is initially prepared in the vacuum state. Note that in the limit $t\gg 1/\xi_\textrm{sq}$ this function asymptotically approaches to $g_{0_\textrm{L},0_\textrm{R}}^{(2)}(t)=2$, and the two modes individually show the statistics of a thermal state.

\section{Further quantum simulation scenarios}\label{sim}
So far we have presented our platform as possible quantum simulator of standard phenomena in quantum optics. However, the versatility of this device suggests further possible uses in other branches of physics, such as in quantum mechanics, multimode quantum systems and quantum thermodynamics. In what follows, we will briefly discuss these three cases.
\subsection{General two-mode coupling for degenerate QHOs}
In quantum mechanics, the Hamiltonian of two interacting bosonic modes induces a linear dynamics when all annihilation and creation operators appear as combinations of quadratic terms \cite{bruschi_general_2021}. These types of Hamiltonian are largely employed in quantum optics, and in this work we already discussed three scenarios of interest. In section \ref{sec:hp}, for example, we discussed a beam splitting interaction, in which the Hamiltonian in Eq.\eqref{hhop} stems from the resonant activation of hopping terms. On the other hand, in section \ref{sec:sq} the Hamiltonian in Eq.\eqref{hsq} only contains the two-mode squeezing terms. In both cases, we made use of the RWA to get rid of highly oscillating terms that do not fulfil the energy conservation. However, the resonance conditions enabling the filtering of the hopping terms and the squeezing terms from Eq.\eqref{intHam} do not exclude themselves reciprocally, but they can be activated together by properly tuning the modulation of the Josephson energy. Once we assume $\tilde\omega=\tilde\upsilon=\omega$ and $\Omega=2\omega$, the interaction Hamiltonian becomes
\begin{align}
\mathcal{\hat H}_\textrm{I}= &\hbar\xi_\textrm{hp} \left(\hat a^\dag\hat b+\hat a\hat b^\dag\right)+\hbar \xi_\textrm{sq}\left(\hat a^\dag\hat b^\dag+\hat a\hat b\right)\nonumber\\
&+\hbar \xi_\textrm{sl}\left[(\hat a^\dag)^2+\hat a^2\right]+\hbar \xi_\textrm{sr}\left[(\hat b^\dag)^2+\hat b^2\right],
\label{hpsq}
\end{align}
where in the second line we introduced the single-mode squeezing coupling terms, weighed by the squeezing parameters for the LHTL and the RHTL mode,
\begin{align}
    \xi_\textrm{sl}=-\frac{\epsilon E_0}{N C_\textrm{l}\omega}\left(\frac{2\pi}{\phi_0}\right)^2,\\
    \xi_\textrm{sr}=-\frac{\epsilon E_0}{N C_\textrm{r}\omega}\left(\frac{2\pi}{\phi_0}\right)^2,
\end{align}
respectively.
This Hamiltonian describes the most general linear interaction between two distinguishable quantum harmonic oscillators (QHOs) characterized by the same frequency.
The dynamics stemming from Eq.\eqref{hpsq} was largely studied by means of the symplectic formalism in \cite{bruschi_general_2021}.
\subsection{Multimode quantum system interaction}
In this section we want to discuss the possibility to extend our model to a larger set of interacting bosonic degrees of freedom. 
We can realize this scenario by shaping the modulation of the Josephson energy as the superposition of $M$ periodic functions:
{\small
\begin{align}
E(t)=E_0\left[1+\sum_{m=1}^{M}(\epsilon_m\sin(\Omega_m t)+\kappa_m\cos(\Omega_m t))\right]
\label{extJE}
\end{align}
with $\epsilon_m$ and $\kappa_m$  oscillation amplitudes, and $\Omega_m$ being the frequency of the $m$-th periodic function. 
}

By choosing the frequencies $\Omega_m$ opportunely, we can therefore activate further resonances and select further interacting modes. This allows us on the one hand to extend the scenarios analyzed so far (beam splitting interactions, Raman scatterings, two-photon squeezing interactions) to a larger number of interacting mode pairs; on the other hand, we can prepare exotic Hamiltonian describing multimode interactions.

As a first example, we can prepare two distinguishable and nondegenerate interacting QHOs \cite{RevModPhys.91.025005}. To do this, we tailor the Josephson energy of the SQUID in Eq.\eqref{extJE} (with $M=2$) by imposing $\Omega_1=\tilde\omega_{h}-\tilde\upsilon$, $\Omega_2=\tilde\omega_{h}+\tilde\upsilon$, $\epsilon_1=\epsilon_2=0$ and $\kappa_1=\kappa_2=2$. Once we set the parameters as above, the interaction Hamiltonian reduces to
\begin{align}
\mathcal{\hat H}_\textrm{I}=\hbar \xi\left(\hat a+\hat a^\dag\right)\left(\hat b+\hat b^\dag\right),
\end{align}
where
\begin{align}
\xi_\textrm{hp}=\xi_\textrm{sq}=\xi=-\frac{\epsilon E_0}{N\sqrt{C_\textrm{r}C_\textrm{l}\omega\upsilon}}\left(\frac{2\pi}{\phi_0}\right)^2.
\label{usp}
\end{align}
The eigenmodes deriving from the diagonalization of this Hamiltonian are sometimes referred to as polaritons \cite{lemonde_nonlinear_2013}, and their dynamics is well known in literature \cite{bruschi_general_2021}.

Interestingly, the possibility to activate more resonances by means of the modulation of the Josephson energy in Eq.\eqref{extJE} permits a drastic extension of the number of interacting modes, and consequently of the Hilbert space. As an example, we can prepare the general Hamiltonian of a three-mode system:
\begin{align}
\mathcal{\hat H}_\textrm{I}=&\hbar \left[\xi_1\left(\hat c_i+\hat c_i^\dag\right)\left(\hat c_j+\hat c_j^\dag\right)+\xi_2\left(\hat c_i+\hat c_i^\dag\right)\left(\hat c_k+\hat c_k^\dag\right)\right.\nonumber\\
&\left.+\xi_3\left(\hat c_j+\hat c_j^\dag\right)\left(\hat c_k+\hat c_k^\dag\right)\right],
\end{align}
with $\xi_1$, $\xi_2$ and $\xi_3$ coupling constants, and with $c=a,b$ representing one mode of either the LHTL or the RHTL. This Hamiltonian describes three selected modes of the CTL, each thereof is interacting with the other one.
\begin{figure}[t]
	\centering
	\includegraphics[width=1\linewidth]{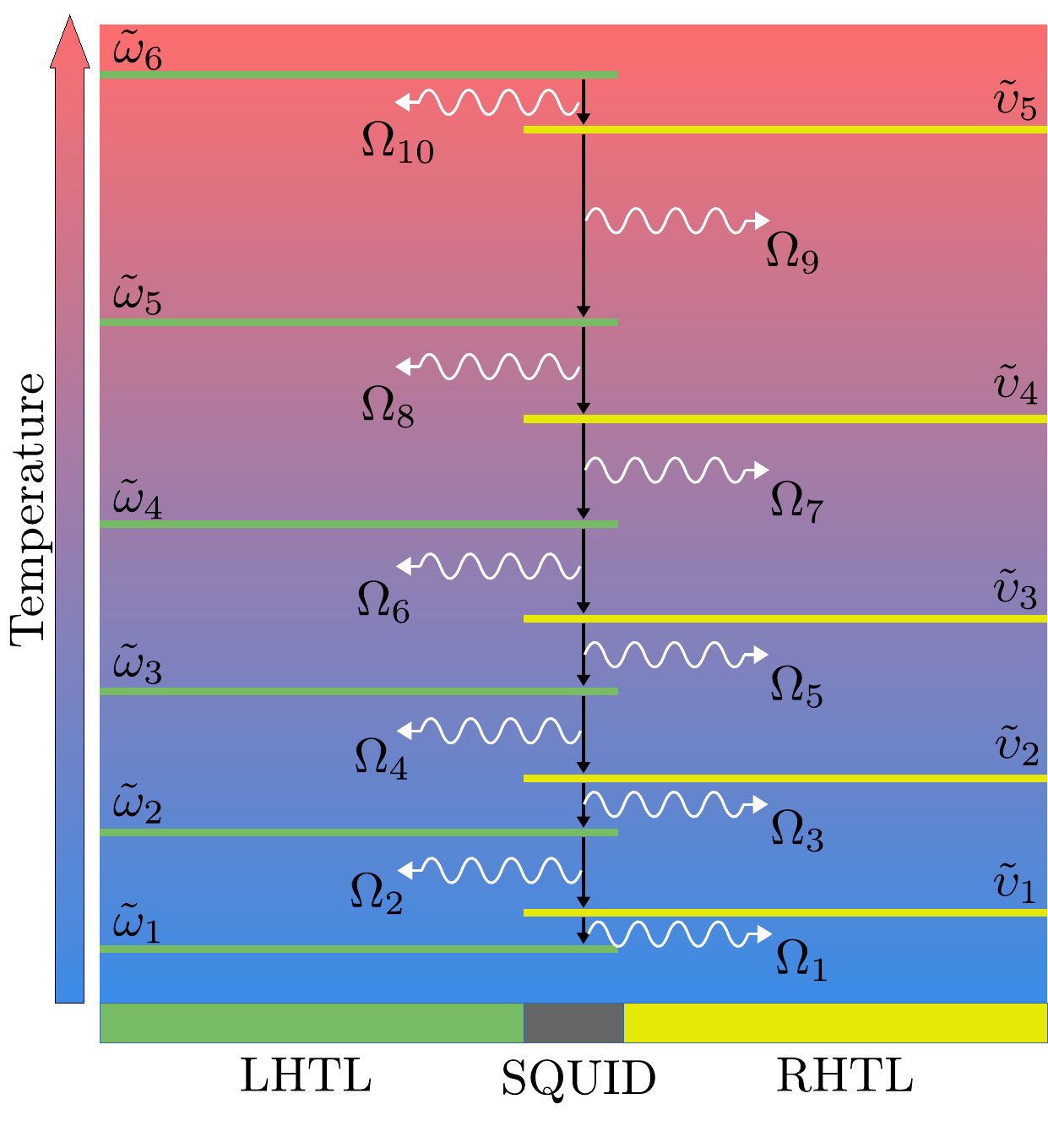}
	\caption{Schematic representation of the cascade quantum network. The modulation frequencies $\Omega_m$ of the Josephson energy are set thereby activating the Raman resonances between modes of the LHTL and the RHTL. Every resonant mode of the CTL is coupled to a bath at temperature $T_{n_\textrm{x}}$, with $\textrm{x}=\textrm{l},\textrm{r}$. Higher frequency modes of the CTL are coupled to baths with higher temperatures. Therefore, the SQUID extracts work at each step of the Raman cascade. The subscripts on the frequencies simply differentiate the mode frequencies and do not refer to any specific modes.}
	\label{network}
\end{figure}
\subsection{CTL as quantum thermal machine}
Finally, we want to discuss some interesting applications of our system in quantum thermodynamics. The idea is to employ the CTL as a quantum thermal machine \cite{quan_maxwells_2006, del_grosso_quantum_2022}, which can either convert heat into work (quantum heat engine) \cite{kosloff_quantum_2014,PhysRevResearch.5.043274} or cool down a cold bath (quantum refrigerator) \cite{kosloff_quantum_2014,levy_quantum_2012}.
A model of quantum heat engine is the quantum amplifier, in which two bosonic modes, coupled to two baths at different temperatures, interact by means of an external drive to generate output power \cite{kosloff_quantum_2014}. To let our device work as a quantum amplifier, we need to thermalize two modes of the CTL, say $\tilde\omega$ and $\tilde\upsilon$, by coupling them to two baths at different temperature: assuming $\tilde\omega>\tilde\upsilon$, the LHTL mode is coupled to a hot bath at temperature $T_\textrm{h}$, whereas the RHTL is coupled to a cold bath at temperature $T_\textrm{c}$. The required resonance condition we need to exploit is the Raman scattering $\Omega=\tilde\omega-\tilde\upsilon$. The interaction Hamiltonian written in the Schrödinger picture is:
\begin{align}
\mathcal{\hat H}_\textrm{I}(t)=\hbar \xi \left(e^{i\Omega t}\hat a\hat b^\dag+e^{-i\Omega t}\hat a^\dag\hat b\right),
\end{align}
where we again assumed a modulation of the Josephson energy of the form 
$E(t)=E_0(1+\epsilon\cos(\Omega t))$. By carrying out the time derivative of this Hamiltonian and averaging it with respect to the state, we get the amplifier output power \cite{kosloff_quantum_2014}:
\begin{align}
\mathcal{P}(t)=i\hbar \Omega\xi \left(e^{i\Omega t}\langle\hat a\hat b^\dag\rangle-e^{-i\Omega t}\langle\hat a^\dag\hat b\rangle\right).
\end{align}

Another powerful application of our CTL is the realization of a quantum network \cite{kosloff_quantum_2013,martinez_dynamics_2013}. 
In the context of quantum thermodynamics, a quantum network consists of an ensemble of wires and junctions which allows the heat propagation and the extraction of net work.
The multimode structure of the CTL, along with the possibility to activate multiple resonances (as shown in the previous section), suggest that the CTL itself can act as a quantum network. 
A model of quantum network can be realized by coupling the modes of the LHTL and the RHTL to different baths and activating chains of Raman resonances by means of a tailored modulation of the Josephson energy. The cascade quantum amplifier is pictorially represented in Fig.\ref{network}. The higher the mode frequency, the higher the temperature of the bath the mode is coupled with. The chain of Raman resonances activates the heat flow from the higher temperature baths towards the cooler temperature ones, whereas the SQUID extracts power at each Raman decay. More details about the implementation of this quantum heat engine on our CTL platform, as well as the analysis of the heat flows and the extracted power, are left for a future publication.

\section{Summary and Conclusion}\label{conc}
We have presented a hybrid platform realized by connecting a left-handed and a right-handed transmission line by means of a superconducting quantum interference device.
After quantizing the magnetic flux fields in both TLs, we have investigated the phase matching at the SQUID, showing that the propagation of quantum excitations from one TL to the other one can occur only for two degenerate modes of the CTL. 

We observed that the interaction between the two TLs occurs only between two modes fulfilling specific resonance conditions. We have shown that we can select the interacting modes, as well as the type of interaction, by controlling the set of circuit parameters. Among all possible interactions we could have activated, we have presented the hopping effect, the Raman scattering, and the two-mode squeezing. Considered more relevant for our discussion, we have focused our attention only on interactions between LHTL and RHTL modes. 

As a result of our analysis, we have shown that our hybrid transmission line can simulate standard processes in quantum mechanics and quantum optics, such as the parametric down-conversion and the HOM interference at the microwave scales. Among other technical uses, our platform can Raman amplify an incoming signal preserving the wave vector, and it can be exploited in the framework of quantum thermodynamics as a quantum heat engine.

As extensions of this work we envisage the realization of chains of hybrid metamaterial TLs in which we can control the spectral feature of the propagating quanta with higher precision. In the framework of quantum metrology, this platform can find application in the realization of circuit versions of well-known interference platforms, such us collinear Mach-Zehnder and SU(1,1) interferometers.
This work paves the way for future theoretical and experimental implementations of hybrid metamaterial transmission lines working at the quantum scales.

\section{Acknowledgments}
The authors thank David Edward Bruschi for the helpful comments. The authors acknowledge support from the joint project No. 13N15685 ``German Quantum Computer based on Superconducting Qubits (GeQCoS)'' sponsored by the German Federal Ministry of Education and Research (BMBF) under the framework “Quantum technologies–from basic research to the market”.
\bibliographystyle{apsrev4-2}
\bibliography{ref}
\appendix

\section{Hamiltonian of hybrid transmission line}\label{app:hyb}
In this section we want to discuss the derivation of our Hamiltonian, as well as the single contribution to the interaction Hamiltonian, in more details. We start from the linearized Lagrangian in Eq.\eqref{LL}, \eqref{LR} and Eq.\eqref{Lint}. By means of the Legendre transformation, 
\begin{align}
\mathcal{H}=\sum_n\left(\frac{\partial \mathcal{L}}{\partial\dot\Phi_n^\textrm{L}}\dot \Phi_n^\textrm{L}+\frac{\partial \mathcal{L}}{\partial\dot\Phi_n^\textrm{R}}\dot \Phi_n^\textrm{R}\right)-\mathcal{L},
\end{align}
we obtain the total Hamiltonian 
\begin{align}
\mathcal{\hat H}=\mathcal{\hat H}_\textrm{L}+\mathcal{\hat H}_\textrm{R}+\mathcal{\hat H}_\textrm{I},\label{hamiltonian}
\end{align}
where
\begin{align}
\mathcal{H}_\textrm{L}=&\frac{1}{2}\sum_{n_\textrm{l}=1}^N\left[\sum_{i=1}^{n_\textrm{l}} \frac{(P_{n_\textrm{l}}^\textrm{L})^2}{C_\textrm{l}}+\frac{(\Phi_{n_\textrm{l}}^\textrm{L})^2}{L_\textrm{l}}\right]\label{HLFP}\\
\mathcal{H}_\textrm{R}=&\frac{1}{2}\sum_{n_\textrm{r}=1}^N\left[C_\textrm{r}(P_{n_\textrm{r}}^\textrm{R})^2+\frac{(\Phi_{n_\textrm{r}+1}^\textrm{R}-\Phi_{n_\textrm{r}}^\textrm{R})^2}{L_\textrm{r}}\right]\label{HRFP}\\
\mathcal{H}_\textrm{I}=&\frac{E(t)}{2}\left(\frac{2\pi}{\phi_0}\right)^2\left(\Phi_0^\textrm{L}-\Phi_0^\textrm{R}\right)^2,\label{HIFP}
\end{align}
with conjugated momenta
\begin{align}
P_{n_\textrm{l}}^\textrm{L}=&\frac{\partial \mathcal{L}^\textrm{L}}{\partial\dot \Phi_{n_\textrm{l}}}=C_\textrm{l}\left(2\dot\Phi_{n_\textrm{l}}^\textrm{L}-\dot\Phi_{{n_\textrm{l}}+1}^\textrm{L}-\dot\Phi_{{n_\textrm{l}}-1}^\textrm{L}\right)\nonumber\\
=&-i\sum_{\lvert j \rvert=1}^{N/2}W_j\left(e^{i(k_j n_\textrm{l}\Delta x-\omega_{j} t)}a_j-\textrm{h.c.}\right),\label{momL}\\
P_{n_\textrm{r}}^\textrm{R}=&\frac{\partial \mathcal{L}^\textrm{R}}{\partial\dot \Phi_{n_\textrm{r}}}=C_\textrm{r}\dot\Phi_{n_\textrm{r}}^\textrm{R}\nonumber\\
=&-i\sum_{\lvert j \rvert=1}^{N/2}\sqrt{\frac{\hbar C_\textrm{r} \upsilon_{j}}{2 N}}\left(e^{i(p_j n_\textrm{r}\Delta x- \upsilon_{j} t)} b_j-\textrm{h.c.}\right).\label{momR}
\end{align}
with $W_j=\sqrt{\frac{8\hbar C_\textrm{l}\omega_{j}}{N}}\sin^2\left(\frac{k_j\Delta x}{2}\right)$, where in the last lines we made use of the mode decomposition of the magnetic flux fields in Eq.\eqref{leftmode} and \eqref{rightmode}.
Substituting the fields and the canonical momenta in Eqs.\eqref{leftmode}, \eqref{rightmode}, \eqref{momL} and \eqref{momR} into the Hamiltonian in Eq.\eqref{hamiltonian}, summing with respect to the spatial indices $n_\textrm{l}$ and $n_\textrm{r}$, and exploiting the commutation relations in Eqs.\eqref{commaad} and \eqref{commbbd}, we obtain the Hamiltonian contributions in Eqs.\eqref{HL}, \eqref{HR} and \eqref{intHamold}. 

Once we expand the square of the interaction Hamiltonian in Eq.\eqref{intHamold}, we distinguish the single contributions to the interaction, as done in Eq.\eqref{intHam}. In what follows, we will present the explicit form of the single terms in Eq.\eqref{intHam} and describe their role in the dynamics.

The first term we want to present is 
\begin{align}
\mathcal{\hat H}_\textrm{ES}=\frac{\hbar E(t)}{N}\left(\frac{2\pi}{\phi_0}\right)^2\sum_{\lvert j \rvert=1}^{N/2}\left(\frac{1}{C_\textrm{l}\omega_j}a_j^\dag\hat a_j+\frac{1}{C_\textrm{r}\upsilon_j}\hat b_j^\dag\hat b_j\right).
\end{align}
It describes the shift of the bare frequencies of the CTL, as discussed in section \ref{encon}. 

The interaction Hamiltonian describes both internal couplings, namely the interactions within the same TL, and external couplings, in which modes from the LHTL and the RHTL interact with each other. We now focus on the internal coupling terms, starting from
\begin{align}
\mathcal{\hat H}_\textrm{RM}=2\chi(t)\sum_{\substack{i,j\\i\neq j}}^{N/2}&\left[\frac{1}{C_\textrm{l}\sqrt{\omega_i\omega_j }}\hat a_i\hat a_j^\dag e^{i(k_i-k_j)\Delta x}\right. \nonumber\\
&\left.+\frac{1}{C_\textrm{r}\sqrt{\upsilon_i\upsilon_j }}\hat b_i\hat b_j^\dag  e^{i(p_i-p_j)\Delta x}\right].
\end{align}
This contribution to the Hamiltonian is related to the internal Raman effect. Indeed, it describes the possibility to annihilate one particle and generate another particle with a different energy, depending on whether the TL absorbs from or releases energy to the SQUID. Clearly, the activation of this effect strongly depends on the time-dependence of the Josephson energy at the SQUID. 

The next two terms include the possibility to squeeze the state of a single TL by tuning the time-dependent Josephson energy. In particular
\begin{align}
\mathcal{\hat H}_\textrm{1S}=\chi(t)\sum_{\lvert j \rvert=1}^{N/2}&\left[\frac{1}{C_\textrm{l}\omega_j}\hat a_j^2e^{2ik_j\Delta x}+\frac{1}{C_\textrm{r}\upsilon_j}\hat b_j^2 e^{2ip_j\Delta x}+\textrm{h.c.}\right]    
\end{align}
describes the squeezing of a single mode of the CTL, generating therefore photon pairs within the same mode, whereas
\begin{align}
\mathcal{\hat H}_\textrm{2S}=\chi(t)\sum_{i,j}^{N/2}&\left[\frac{1}{C_\textrm{l}\sqrt{\omega_i\omega_j }}\hat a_i\hat a_j e^{i(k_i+k_j)\Delta x}\right. \nonumber\\
&\left.+\frac{1}{C_\textrm{r}\sqrt{\upsilon_i\upsilon_j }}\hat b_i\hat b_j e^{i(p_i+p_j)\Delta x}+\textrm{h.c.}\right],  
\end{align}
describes the two-mode squeezing, namely the creation of two photons in two different modes of the same TL.

Finally, we present the two external coupling terms, which have important implications in the interaction between the LHTL and the RHTL, and for this reason are largely discussed in the main text. As first, we introduce the hopping effect, associated with the Hamiltonian contribution
\begin{align}
\mathcal{\hat H}_\textrm{HP}=&\frac{\chi(t)}{\sqrt{C_\textrm{l}C_\textrm{r}}}\sum_{i,j}^{N/2}\frac{1}{\sqrt{\omega_i\upsilon_j}}\left(\hat a_i\hat b_j^\dag e^{i(k_i-p_j)\Delta x}+\textrm{h.c.}\right).
\end{align}
This term allows the propagation of excitations form one TL to the other one. Thank to the time-dependence of the Josephson energy, the hopping effect can occur also between modes having different energies. In this case, this effect corresponds to an external Raman scattering.

To conclude, the last contribution to the Hamiltonian is
\begin{align}
\mathcal{\hat H}_\textrm{IS}=-\frac{\hbar E(t)}{N\sqrt{C_\textrm{l}C_\textrm{r}}}\left(\frac{2\pi}{\phi_0}\right)^2\sum_{i,j}^{N/2}\frac{1}{\sqrt{\omega_i\upsilon_j}}\left(a_i^\dag\hat b_j^\dag +\hat a_i\hat b_j\right).
\end{align}
This term implies the possibility to squeeze two modes stemming from two different TL. 
\end{document}